\documentclass[aps,prl,twocolumn,superscriptaddress,nofootinbib]{revtex4-2}

\usepackage{amsmath,amssymb,graphicx,bm}
\usepackage{hyperref}
\usepackage{tikz}
\usepackage{pgfplots}
\pgfplotsset{compat=1.18}

\begin{document}

\title{Probing Freeze-In Dark Matter via a Spin-2 Portal at the LHC with Vector Boson Fusion and Machine Learning}

\author{Junzhe Liu$^1$, Alfredo Gurrola}
\affiliation{Department of Physics and Astronomy, Vanderbilt University, Nashville, TN, USA}

\begin{abstract}
The persistent absence of signals in traditional dark matter searches has intensified interest in scenarios beyond the canonical weakly interacting massive particle paradigm. In this work, we investigate the collider phenomenology of feebly interacting dark matter produced via the freeze-in mechanism through a spin-2 portal. We consider a framework in which a massive graviton-like mediator couples minimally and universally to the energy--momentum tensor of both the Standard Model (SM) and the dark sector. Such interactions arise naturally in extra-dimensional constructions and effective theories of gravity, providing a theoretically well-motivated and predictive setup. We systematically connect early-Universe cosmology with collider observables by identifying regions of parameter space consistent with freeze-in conditions and the observed dark matter relic abundance, and examining their testability at the Large Hadron Collider (LHC). Focusing on bosonic fusion production channels, which are particularly sensitive to spin-2 interactions, we analyze invisible mediator decay signatures and assess current and projected experimental sensitivities. To enhance sensitivity in this challenging regime of feeble couplings, we develop a search strategy based on machine-learning algorithms. Our results demonstrate that collider searches can probe substantial regions of the cosmologically viable freeze-in parameter space, highlighting the high-luminosity LHC as a powerful laboratory for feebly interacting dark sectors. This study establishes a concrete and complementary pathway to test freeze-in dark matter scenarios through spin-2 portals, thereby bridging gravitationally motivated new physics, cosmology, and high-energy collider experiments.
\end{abstract}

\maketitle

\section{Introduction}
\label{introduction} 
The nature of dark matter (DM) remains one of the most compelling mysteries in both particle physics and cosmology. While astrophysical and cosmological observations provide overwhelming evidence for the existence of DM~\cite{WMAP:2012nax,Planck:2018vyg,Bertone:2004pz}, its particle nature is still unknown. The canonical weakly interacting massive particle (WIMP) paradigm has long dominated the theoretical landscape~\cite{Feng:2022inv,Arcadi:2017kky,Florez:2016lwi,Dutta:2008ge,Arnowitt:2008bz}; however, the absence of conclusive signals from direct detection and collider experiments~\cite{Arcadi:2017kky,Arcadi:2024waning} has motivated growing interest in alternative frameworks, including models with \textit{feebly interacting dark matter} (FIDM)~\cite{Bernal:2017kxu,Junius:2022fidm}. In these scenarios, the DM particles interact so weakly with the Standard Model (SM) that they have a different early universe cosmology, evade conventional searches, necessitating novel theoretical and experimental approaches.

Feebly interacting dark matter can arise naturally in theories beyond the Standard Model (BSM), particularly those that postulate new portals between the SM and a hidden or dark sector. A compelling class of such portals involves the exchange of spin-2 mediators, massive graviton-like particles, which couple to the energy-momentum tensor of both visible and dark sectors~\cite{Han:1998sg,Giudice:1998ck}. These mediators can emerge in extra-dimensional models, such as Randall--Sundrum scenarios~\cite{Randall:1999ee,Randall:1999vf}, or in low-energy effective theories of quantum gravity~\cite{Donoghue:1994dn,Burgess:2003jk}. Notably, spin-2 portals offer a universal coupling structure, making them theoretically appealing and comparatively constrained.

From a cosmological standpoint, FIDM models often realize the \textit{freeze-in} mechanism for DM production. Unlike the thermal freeze-out mechanism characteristic of WIMPs, freeze-in production involves the gradual accumulation of dark matter particles from rare SM interactions 
wherein DM particles possess feeble interactions with the SM sector and never attain thermal equilibrium in the early Universe~\cite{Bernal:2017kxu,Hall:2009bx,Elahi:2014fsa}. Instead, their relic abundance is generated gradually through out-of-equilibrium processes, resulting in distinct phenomenology and parameter space compared to thermal freeze-out models. 
In this context, the production rate is typically governed by the mass and couplings of the mediator, making collider experiments sensitive probes of the underlying dynamics. Crucially, if the mediator is within the kinematic reach of current or future colliders, such as the Large Hadron Collider (LHC) or proposed next-generation facilities, it may be produced on-shell and decay visibly or invisibly, offering distinctive experimental signatures.

This work explores the potential of collider experiments to probe FIDM via spin-2 mediators. We focus on theoretical models where the mediator couples minimally to the SM through the energy-momentum tensor and is responsible for the production of DM in the early Universe via freeze-in. We aim to bridge the gap between early-Universe cosmology and collider physics, identifying viable parameter regions consistent with cosmological relic abundance, and exploring the discovery prospects and constraints achievable through novel search methods utilizing bosonic fusion processes at the high-luminosity LHC (HL-LHC).

\section{Theoretical Framework: A Spin-2 Portal to Scalar Dark Matter}

We consider an extension of the SM in which the DM particle resides in a hidden sector and communicates with the visible sector exclusively through a massive spin-2 mediator, denoted $G_{\mu\nu}$. The field $G_{\mu\nu}$ is taken to be a symmetric rank-2 tensor describing a massive spin-2 state, analogous to a Kaluza--Klein graviton or a composite tensor resonance~\cite{Han:1998sg,Giudice:1998ck,Contino:2001si,Fierz:1939ix}. Consistency with Lorentz invariance and the absence of ghost-like degrees of freedom at the linear level motivate couplings of $G_{\mu\nu}$ to conserved energy--momentum tensors. Accordingly, we construct the interaction Lagrangian in terms of lowest-dimension effective operators that couple $G_{\mu\nu}$ to the energy--momentum tensors of the visible and dark sectors.

In this section, we focus on a minimal realization in which the spin-2 mediator couples to SM photons and to a scalar dark matter field, but not to other SM degrees of freedom. This defines a ``photon-only portal'' between the two sectors. While such a restriction is not mandatory in a generic ultraviolet (UV) completion, it provides a clean and predictive benchmark that isolates the role of electromagnetic initial states in both cosmology and collider phenomenology. In later sections, we extend the discussion to more general scenarios in which the spin-2 mediator also couples to the full electroweak gauge sector of the SM, thereby assessing the  conclusions beyond the photon-only limit and exploring the impact of additional production and decay channels.

The DM candidate is assumed to be a real, stable scalar field $\chi$, singlet under the SM gauge group. Stability may be ensured, for instance, by imposing a discrete $\mathbb{Z}_2$ symmetry under which $\chi \to -\chi$. The interaction Lagrangian is given by
\begin{equation}
\mathcal{L}_{\text{int}} 
= \frac{1}{\Lambda_\gamma} \, G^{\mu\nu} T^{(\gamma)}_{\mu\nu}
+ \frac{1}{\Lambda_\chi} \, G^{\mu\nu} T^{(\chi)}_{\mu\nu},
\end{equation}
where $\Lambda_\gamma$ and $\Lambda_\chi$ are independent mass scales controlling the strength of the mediator couplings to photons and to dark matter, respectively. These parameters characterize the effective cutoff scales of the theory and encode the underlying UV dynamics, such as the compactification scale in extra-dimensional constructions or the compositeness scale in strongly coupled realizations~\cite{Giudice:1998ck,Randall:1999ee,ArkaniHamed:1998rs,Contino:2010rs,Panico:2015jxa}. Allowing $\Lambda_\gamma \neq \Lambda_\chi$ captures the possibility of asymmetric couplings between the visible and dark sectors, which is well motivated in scenarios where the two sectors have distinct localizations or origins in the UV completion.

The relevant energy--momentum tensors are
\begin{align}
T_{\mu\nu}^{(\gamma)} &= F_{\mu\alpha}F_{\nu}^{\;\alpha} 
- \frac{1}{4} \eta_{\mu\nu} F^{\alpha\beta} F_{\alpha\beta}, \\
T_{\mu\nu}^{(\chi)} &= \partial_\mu \chi \, \partial_\nu \chi 
- \eta_{\mu\nu} \left( \frac{1}{2} \partial^\alpha \chi \, \partial_\alpha \chi 
- \frac{1}{2} m_\chi^2 \chi^2 \right),
\end{align}
where $F_{\mu\nu}$ is the electromagnetic field strength tensor and $m_\chi$ is the scalar DM mass. The form of these tensors ensures gauge invariance and conservation of energy and momentum in the limit of vanishing mediator mass, while the resulting operators are dimension five and therefore suppressed by the high scales $\Lambda_\gamma$ and $\Lambda_\chi$. For sufficiently large values of these scales, the interactions are naturally feeble, as required in freeze-in scenarios.

This construction yields a particularly clean portal structure in which $G_{\mu\nu}$ is the sole mediator between the SM and dark sectors, and no direct renormalizable interactions between $\chi$ and SM fields are introduced. By design, the mediator does not couple at tree level to SM fermions or to gluons. As a consequence, production mechanisms at hadron colliders that rely on quark or gluon initial states are absent at leading order, substantially suppressing conventional Drell--Yan, dijet, and dilepton resonance signatures that typically provide strong constraints on generic spin-2 or graviton-like states. Any effective coupling to charged fermions is induced only radiatively through photon loops and is therefore additionally suppressed by loop factors and by the large scale $\Lambda_\gamma$. 

The absence of direct couplings to electrons, nucleons, and gluons also significantly weakens astrophysical and direct detection bounds~\cite{Essig:2013lka,Alexander:2016aln}. In particular, elastic $\chi$--nucleon scattering proceeds only at higher order and is highly suppressed, allowing the model to evade stringent limits from underground direct detection experiments~\cite{Aprile:2018dbl,Akerib:2016vxi}. Similarly, mediator production in stellar interiors through processes involving electrons or nucleons is strongly reduced, alleviating constraints from stellar cooling and supernova energy-loss arguments that often exclude light, weakly coupled mediators~\cite{Raffelt:1996wa,Hardy:2016kme}. 

In the early Universe, the dominant DM production channel is photon--photon fusion,
\begin{equation}
\gamma\gamma \rightarrow G^*/G \rightarrow \chi\chi,
\end{equation}
with the rate controlled by $\Lambda_\gamma$, $\Lambda_\chi$, and the mediator mass $m_G$. For sufficiently large cutoff scales, the interaction rate remains below the Hubble expansion rate at all times, preventing thermal equilibration of $\chi$ and realizing the freeze-in mechanism. The framework is therefore characterized by a small and predictive parameter set: the mediator mass $m_G$, the dark matter mass $m_\chi$, and the effective scales $\Lambda_\gamma$ and $\Lambda_\chi$. In the following sections, we compute the freeze-in yield from $\gamma\gamma \to \chi\chi$, identify the regions of parameter space consistent with the observed relic abundance, and analyze the corresponding collider phenomenology associated with spin-2 mediator production and decay at the LHC.

\section{Freeze-in Conditions and Relic Density Calculation}

A DM candidate is said to be in thermal equilibrium with the SM bath if its interaction rate is sufficiently large to maintain a thermal phase-space distribution with the same temperature as the visible plasma. Quantitatively, this requires that the interaction rate per particle, $\Gamma(T) \sim n_{\rm bath}(T)\langle\sigma v\rangle$, exceeds the Hubble expansion rate $H(T)$. Here $n_{\rm bath}$ denotes the number density of bath particles participating in the interaction and $\langle\sigma v\rangle$ is the thermally averaged cross section. In this regime, number-changing processes occur rapidly compared to the expansion of the Universe, and the DM number density tracks its equilibrium value $n_\chi^{\rm eq}$.

In contrast, the freeze-in mechanism assumes that the couplings between the dark sector and the SM are extremely small. As a consequence, $\Gamma(T) < H(T)$ for all relevant temperatures $T$, so that the DM particle never attains thermal equilibrium~\cite{Hall:2009bx}. Its abundance is instead gradually generated through rare scatterings or decays of particles in the thermal bath. Since $\Gamma < H$, inverse processes that would deplete the DM population are negligible, and the Boltzmann equation simplifies considerably.

Throughout this work we assume a standard radiation-dominated cosmological history after reheating, with entropy conservation in the visible sector and no significant late-time entropy injection. 
%We further assume that the reheating temperature $T_R$ is larger than the relevant mass scales in the problem, unless otherwise stated. 
Under these assumptions, the Hubble parameter and entropy density are given by
\begin{equation}
    H(T) = 1.66 \sqrt{g_*} \frac{T^2}{M_{\rm Pl}}, 
    \qquad
    s(T) = \frac{2\pi^2}{45} g_{*s} T^3,
\end{equation}
where $M_{\rm Pl} = 1.22\times10^{19}\,$GeV is the Planck mass and $g_*$ and $g_{*s}$ denote the effective numbers of relativistic degrees of freedom contributing to the energy and entropy densities, respectively. 

\begin{figure*}
    \centering
    \begin{minipage}{.5\textwidth}
  \centering
  \includegraphics[width=\linewidth]{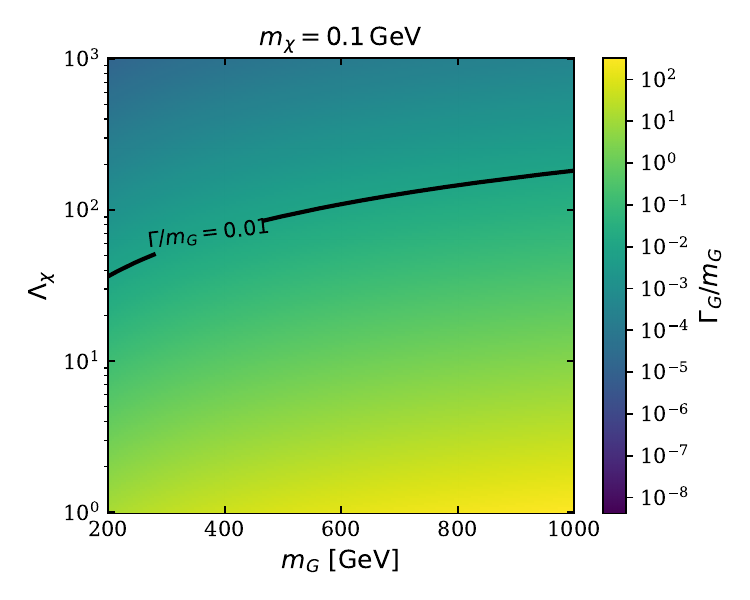}
\end{minipage}%
\begin{minipage}{.5\textwidth}
  \centering
  \includegraphics[width=\linewidth]{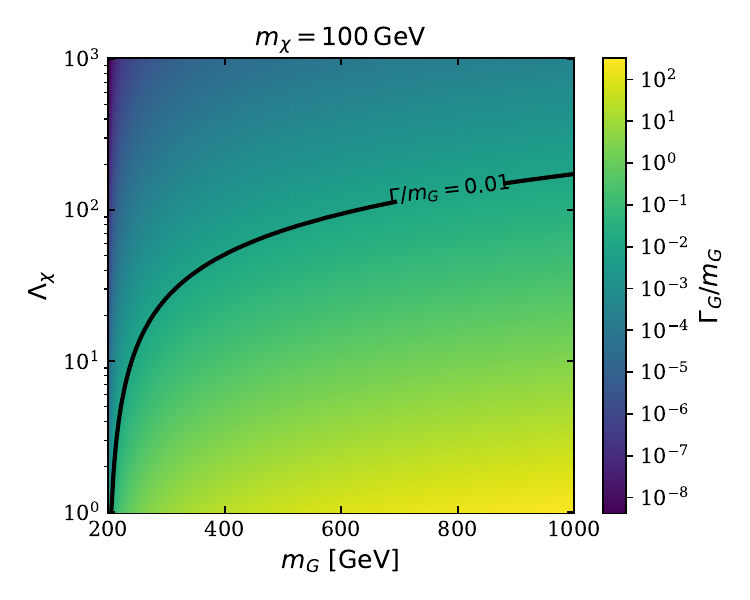}
\end{minipage}%
    \caption{Ratio of decay width to spin-2 mediator mass for dark matter masses of $m_\chi = 0.1~\text{GeV}$ and $m_\chi = 100~\text{GeV}$. The solid line represents the contour of constant relative decay width $\Gamma / m_G = 0.01$.}
    \label{fig:decayWidth}
\end{figure*}

It is convenient to express the Boltzmann equation in terms of the comoving yield $Y \equiv n_\chi/s$. Neglecting back-reaction terms (valid in the freeze-in regime), one obtains
\begin{equation}
    \frac{dY}{dT}
    = -\,\frac{C(T)}{s(T)\,H(T)\,T},
\end{equation}
where $C(T)$ is the collision term computed using equilibrium distributions for the SM bath particles~\cite{Bernal:2017kxu,Hall:2009bx}. In our spin-2 portal framework, the dominant production process is photon--photon fusion, $\gamma\gamma \rightarrow G^*/G \rightarrow \chi\chi$. The collision term can be written schematically as
\begin{equation}
\begin{split}
    C(T) = \int 
    (2\pi)^4 &\delta^{(4)}(p_i - p_f)
    |\mathcal{M}|^2 
    f_\gamma^{\rm eq} f_\gamma^{\rm eq} \times \\
    &d\Pi_\gamma d\Pi_\gamma d\Pi_\chi d\Pi_\chi 
\end{split}
\end{equation}
where $d\Pi_i$ denotes the Lorentz-invariant phase space measure and $f_\gamma^{\rm eq}$ are the equilibrium Bose--Einstein distributions. The squared matrix element $|\mathcal{M}|^2$ is determined by the spin-2 propagator and the effective couplings $1/\Lambda_\gamma$ and $1/\Lambda_\chi$. Parametrically, away from resonance one finds
\begin{equation}
    \langle\sigma v\rangle \;\propto\;
    \frac{T^6}{\Lambda_\gamma^2 \Lambda_\chi^2\, m_G^4},
\end{equation}
reflecting the higher-dimensional nature of the interaction and the derivative structure of the energy--momentum tensor couplings. Consequently, the production rate grows rapidly with temperature. Integrating the Boltzmann equation from $T=0$ to $T=T_R$, one obtains
\begin{equation}
    Y(T_R) 
    = \int_0^{T_R} 
    \frac{C(T)}{s(T)\,H(T)\,T}\, dT.
\end{equation}
If the integrand increases with temperature, the integral is dominated by the upper limit $T_R$, leading to so-called ultraviolet freeze-in~\cite{Elahi:2014fsa}. In this case, the final yield depends explicitly on the reheating temperature. For the off-resonant regime, one finds parametrically
\begin{equation}
    Y_{\chi} 
    \;\simeq\; 
    \kappa\,
    \frac{M_{\rm Pl}}{g_*^{3/2}}
    \frac{T_R^{7}}
    {\Lambda_\gamma^2 \Lambda_\chi^2\, m_G^4},
\end{equation}
where $\kappa$ is a numerical coefficient obtained from the phase-space integration. The strong $T_R^7$ dependence arises from the combination of the temperature-scaling of the cross section and the cosmological prefactors in the Boltzmann equation.

A practical and conservative criterion for freeze-in is that, at all temperatures, the production rate is too small to bring the DM into equilibrium. This may be expressed as
\begin{equation}
    \frac{C(T)}{H(T)\,n_\chi^{\rm eq}(T)} \ll 1,
    \qquad
    n_\chi^{\rm eq}(T) \simeq \frac{\zeta(3)}{\pi^2}T^3.
\end{equation}
In the off-resonant regime ($m_G \gg T$ during production), this condition is typically most stringent at $T \simeq T_R$, since the production rate grows with temperature. The resulting bound translates into constraints on the combination $\Lambda_\gamma \Lambda_\chi$ for given $m_G$ and $T_R$:
\begin{equation}
(\Lambda_{\gamma}\Lambda_{\varphi})^2 \gg 92.94\;\frac{M_{\rm Pl}}{\sqrt{g_{*}}}\;\frac{T_R^7}{m_G^4}.
\label{eq:freezeInCondition}
\end{equation}

On the other hand, in the resonant regime, where $T \sim m_G$ and the mediator can be produced on-shell, the production is dominated by the Breit--Wigner enhancement of the propagator~\cite{Bernal:2017kxu,Garny:2017rxs}. In this case, a conservative freeze-in requirement is that the mediator itself never thermalizes. This can be imposed through $\Gamma_G \ll H(T \sim m_G)$, where $\Gamma_G$ is the total decay width of the spin-2 mediator, including decays into photons and DM. Figure~\ref{fig:decayWidth} shows the ratio of the decay width to the spin-2 mediator mass, for DM masses of 0.1 GeV and 100 GeV. Since
\begin{equation}
    \Gamma_G = 
    \frac{m_G^3}{80\pi \Lambda_\gamma^2}
    + \frac{m_G^3}{960\pi \Lambda_\chi^2}\left(1-\frac{4m_\chi^2}{m_G^2}\right)^{5/2},
\label{eq:DecayWidth}
\end{equation}
the requirement $\Gamma_G \ll H(m_G)$ places an upper bound on the effective couplings and ensures that $G_{\mu\nu}$ does not enter equilibrium with the plasma. In this regime, the final yield is largely determined by production near $T \sim m_G$ and exhibits a milder dependence on $T_R$.

Once the comoving yield has been determined, the present-day relic abundance follows from
\begin{equation}
    \Omega_\chi h^2 
    = \frac{m_\chi\, s_0\, Y_\chi}{\rho_c/h^2},
\end{equation}
where the current entropy density is $s_0 \simeq 2890 \;{\rm cm}^{-3}$ and the critical density is $\rho_c \simeq 1.05 \times 10^{-5}\, h^2 \, {\rm GeV\,cm}^{-3}$. Numerically, this can be written as
\begin{equation}
    \Omega_\chi h^2 
    \simeq 2.75\times10^8
    \left(\frac{m_\chi}{\rm GeV}\right)
    Y_\chi.
    \label{eq:relicDensity}
\end{equation}

\begin{figure*}
    \centering
    \begin{minipage}{.5\textwidth}
  \centering
  \includegraphics[width=\linewidth]{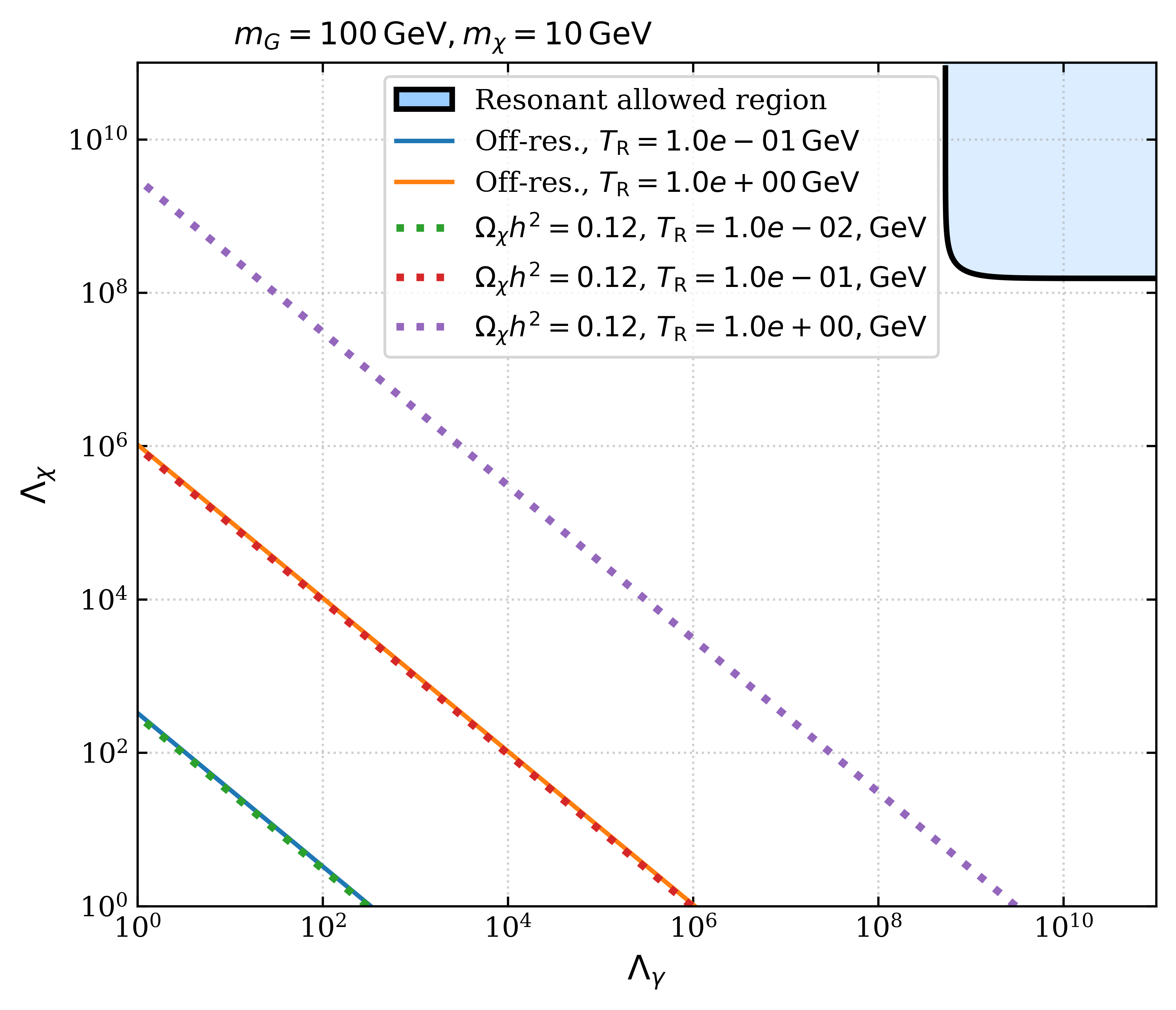}
\end{minipage}%
\begin{minipage}{.5\textwidth}
  \centering
  \includegraphics[width=\linewidth]{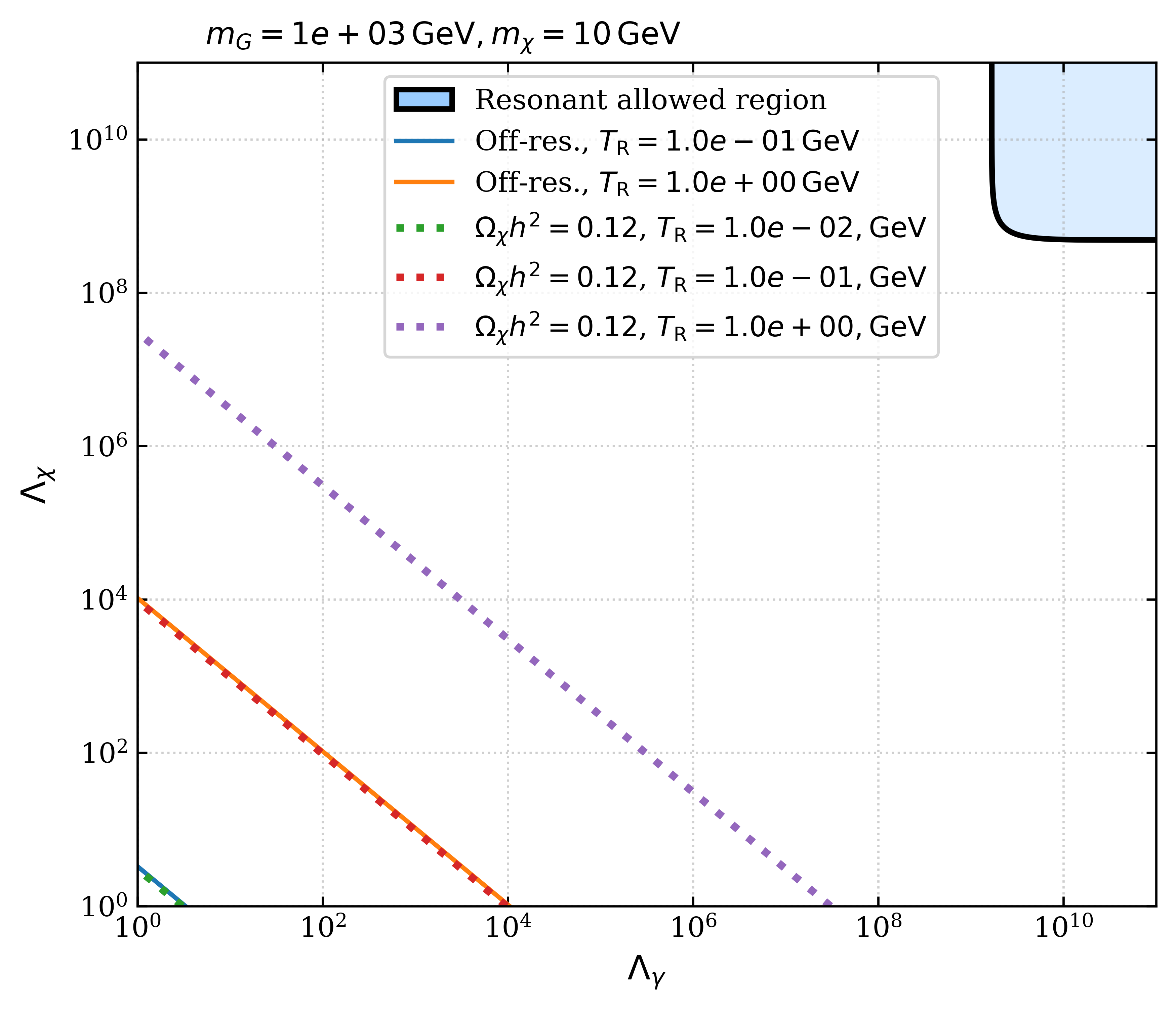}
\end{minipage}%
    \caption{Theoretical constraints on the mediator couplings to dark matter and photons, derived from resonant and off-resonant freeze-in production, as well as the requirement that the relic abundance satisfies $\Omega_\chi h^2 \leq 0.12$ for reheating temperatures of $10~\mathrm{MeV}$, $100~\mathrm{MeV}$, and $1~\mathrm{GeV}$. For the line contours, the region above each line denotes the allowed parameter space.}
    \label{fig:CosmoAllowed}
\end{figure*}

A key qualitative difference from the conventional thermal freeze-out scenario is the dependence of the relic density on the interaction strength. In freeze-out, $\Omega_\chi h^2$ is inversely proportional to the annihilation cross section, $\Omega_\chi h^2 \propto 1/\langle\sigma v\rangle$, since stronger interactions keep DM in equilibrium longer and reduce the final abundance. In freeze-in, by contrast, the relic density is directly proportional to the (squared) feeble couplings, as production is controlled by rare processes that are never balanced by inverse reactions. Consequently, larger values of $\Lambda_\gamma$ and $\Lambda_\chi$ suppress the yield, while higher reheating temperatures can significantly enhance it. This distinct parametric behavior underlies the strong interplay between cosmology and collider physics in spin-2 portal freeze-in scenarios.

Figure~\ref{fig:CosmoAllowed} illustrates the regions of parameter space consistent with cosmological requirements in the spin-2 portal FIDM scenario for two representative mass choices, $\{ m_G, m_{\chi} \} = \{ 100~\rm{GeV}, 10~\rm{GeV} \}$ and $\{ 1000~\rm{GeV},\allowbreak 10~\rm{GeV} \}$, respectively. The solid curves indicate the parameter space satisfying the off-resonant freeze-in condition, corresponding to the regime in which mediator production occurs far from the pole ($m_G \gg T$), as defined by Eq.~\ref{eq:freezeInCondition}. The blue shaded region denotes the parameter space consistent with the resonant freeze-in requirement that the mediator itself never thermalizes with the plasma, $\Gamma_G < H(T \sim m_G)$, as given by Eq.~\ref{eq:DecayWidth}. 

For the off-resonant regime, we consider several representative reheating temperatures $T_R$ in the MeV--GeV range. This range is motivated by cosmological considerations, since successful Big Bang nucleosynthesis requires $T_R \gtrsim \mathcal{O}({\rm MeV})$~\cite{Kawasaki:1999na,Hannestad:2004px,deSalas:2015glj}. In addition to the freeze-in conditions, the figures also display the parameter regions compatible with the observed DM relic abundance. These are shown by the dashed curves obtained by requiring that the relic density computed using Eq.~\ref{eq:relicDensity} matches the measured value $\Omega_\chi h^2 \simeq 0.12$. The intersection of the freeze-in conditions with the relic-density requirement therefore defines the cosmologically viable phase space of the model.

One observes that for reheating temperatures in the MeV--GeV range, where off-resonant production dominates, the allowed parameter space corresponds to effective coupling scales $\Lambda_\gamma$ and $\Lambda_\chi$ typically ranging from the electroweak scale up to the multi-TeV regime. In contrast, for higher reheating temperatures where production near the mediator pole becomes important and the resonant regime is realized, the viable parameter space shifts dramatically toward much larger coupling scales, $\Lambda_\gamma,\,\Lambda_\chi \sim 10^8$--$10^9$~GeV. As we will show in the following sections, the lower-$T_R$ regime associated with off-resonant freeze-in is particularly interesting from an experimental standpoint, since the corresponding parameter space falls within the sensitivity reach of the HL-LHC.

\section{Experimental Considerations}\label{sec:exp}

The most stringent constraints on our model of FIDM arise from LHC searches for high-mass diphoton resonances. Both ATLAS and CMS have performed dedicated searches for narrow spin-2 resonances, often interpreted in the context of Randall--Sundrum (RS) gravitons, using the full Run~2 dataset at $\sqrt{s}=13$~TeV~\cite{ATLAS:2021uiz,CMS:2024diphoton}. In particular, we make use of the published ATLAS results of Ref.~\cite{ATLAS:2021uiz} and the recent CMS analysis of Ref.~\cite{CMS:2024diphoton}, which provide observed and expected $95\%$ confidence level (CL) upper limits on the signal production cross section times branching fraction, $\sigma(pp \to G)\times \mathcal{B}(G\to\gamma\gamma)$, as a function of the resonance mass.

Using the information provided in the ATLAS and CMS diphoton resonance analyses, including details on signal acceptance as well as photon reconstruction and identification efficiencies, we first validate our analysis framework by reproducing the expected upper limits on the signal cross section reported by the experiments. We find agreement at the level of better than $10\%$, demonstrating that our implementation of the signal simulation, detector effects, and event selection faithfully captures the performance of the experimental analyses.

Having established this validation, we then reinterpret the analyses within the context of our spin-2 portal FIDM model by applying the same signal region selections to simulated $pp \to G + X \to \gamma\gamma + X$ events. In particular, we evaluate the signal acceptance under the experimental selection criteria, which are largely driven by requirements on the diphoton invariant mass and photon transverse energy. Since the diphoton invariant mass is a boost-invariant observable and serves as the primary discriminant in these searches, and because the photon transverse energy thresholds employed in the analyses are relatively modest compared to the characteristic scale set by $m_G$, we find that the signal acceptance in our model closely tracks that of the benchmark scenarios used by ATLAS and CMS. Quantitatively, the acceptance differs by no more than $\mathcal{O}(20\%)$ across the parameter space considered.

This level of agreement indicates that the published experimental limits on $\sigma \times \mathcal{B}(G\to\gamma\gamma)$ can be reliably reinterpreted in our framework with minimal model dependence. Consequently, the existing ATLAS and CMS results provide a robust basis for possibly constraining the spin-2 portal parameter space, even though the underlying production mechanism in our scenario is dominated by photon fusion rather than gluon fusion. 

Therefore, given the above considerations, to recast the LHC results in the context of our spin-2 portal model, we proceed as follows. For each mediator mass $m_G$, the experimental analyses provide an upper bound on the signal production cross section times branching fraction, which we denote as $\left[\sigma(pp \to G+X)\times \mathcal{B}(G\to\gamma\gamma)\right]_{\rm LHC}^{\rm{95\%} \rm{ CL}}$.
We then compute the corresponding theoretical prediction within our framework and compare it to the experimental limit.

In our photon-only portal scenario, the dominant production mechanism is photon fusion. The inclusive production cross section scales with the effective coupling as $\sigma(pp \to G+X) \propto \Lambda_\gamma^{-2}$, since the amplitude for $\gamma\gamma\to G$ is proportional to $1/\Lambda_\gamma$. The total diphoton signal rate is therefore given by
\begin{equation}
\sigma(pp \to G \to \gamma\gamma)
= \sigma(pp \to G)\times \mathcal{B}(G\to\gamma\gamma).
\end{equation}
The branching ratio depends sensitively on the relative strength of the couplings to photons and dark matter. For a mediator that can decay into both $\gamma\gamma$ and $\chi\chi$, the diphoton branching fraction is
\begin{equation}
\mathcal{B}(G\to\gamma\gamma)
=
\frac{1}{
1+\dfrac{1}{12}\,\dfrac{\Lambda_\gamma^2}{\Lambda_\chi^2}
\left(1-\dfrac{4m_\chi^2}{m_G^2}\right)^{5/2}
},
\end{equation}
where the second term in the denominator accounts for the partial width into dark matter, including the characteristic $(1-4m_\chi^2/m_G^2)^{5/2}$ phase-space suppression factor for a spin-2 decay into scalars. In the limit $m_G < 2m_\chi$, the invisible channel is kinematically closed and $\mathcal{B}(G\to\gamma\gamma)\to 1$, maximizing the sensitivity of diphoton searches. Conversely, when $m_G > 2m_\chi$ and $\Lambda_\chi \ll \Lambda_\gamma$, the invisible decay width dominates and the diphoton branching ratio is suppressed, weakening the corresponding constraints.

For each point in the parameter space $(m_G,\Lambda_\gamma,\Lambda_\chi,\allowbreak m_\chi)$, we compute the inclusive production cross section $\sigma(pp\to G+X)$ using photon parton distribution functions and multiply by the corresponding branching ratio $\mathcal{B}(G\to\gamma\gamma)$. The resulting prediction is then directly compared to the experimental upper limits from ATLAS and CMS. Parameter points for which $\sigma(pp \to G \to \gamma\gamma) 
>
\left[\sigma\times\mathcal{B}\right]_{\rm LHC}^{95\%{\rm CL}}$ are ruled out at $95\%$~CL. As a reference, current ATLAS and CMS diphoton resonance searches report upper limits of $\left[\sigma\times\mathcal{B}\right]_{\rm LHC}^{95\%{\rm CL}} \sim \mathcal{O}(3)\,\text{fb}$ at $m_G \simeq 500~\text{GeV}$ and $\sim \mathcal{O}(0.2)\,\text{fb}$ at $m_G \simeq 1000~\text{GeV}$~\cite{ATLAS:2021uiz,CMS:2024diphoton}. In contrast, within the spin-2 portal framework considered here, the predicted signal cross section times branching fraction remains below $\sim 1\,\text{fb}$ for $m_G = 500~\text{GeV}$ and below $\sim 0.05\,\text{fb}$ for $m_G = 1000~\text{GeV}$ across the full range of $\Lambda_\gamma$ and $\Lambda_\chi$ values studied. Consequently, no regions of parameter space are currently excluded by existing inclusive diphoton resonance searches.

It is important to emphasize that these experimental analyses are primarily optimized for gluon-fusion production of RS gravitons and rely on high-$p_T$ photon selection strategies. In contrast, our scenario is dominated by photon-fusion production, which leads to smaller cross sections due to the suppressed photon parton luminosity, as well as kinematic features that are not optimally captured by standard search strategies. Moreover, existing analyses typically lose sensitivity for mediator masses below $m_G \sim 500~\text{GeV}$ due to trigger and selection requirements, further limiting their applicability to the parameter space of interest. As a result, although current searches exclude spin-2 resonances up to masses of order $4$--$5$~TeV in strongly coupled scenarios, they provide no meaningful constraints on the feebly coupled photon-fusion-dominated regime considered in this work. This underscores the need for dedicated search strategies tailored to FIDM models, particularly those leveraging advanced machine-learning techniques to enhance sensitivity in regimes with suppressed production rates and challenging signal-to-background discrimination. In the following sections, we develop such an approach and assess its projected impact at the HL-LHC.

\section{Samples and Simulation}
Event generation for both signal and background processes is performed within a Monte Carlo framework based on \verb|FeynRules|~\cite{Alloul:2013bka}. The BSM Lagrangian introduced in the previous sections is implemented in \verb|FeynRules| and exported in the Universal FeynRules Output (UFO) format~\cite{Degrande:2011ua}. These UFO model files are interfaced with \verb|MadGraph5_aMC@NLO v3.5.6|~\cite{Alwall:2014hca,Frederix:2018nkq} for matrix-element generation. All samples are produced using the NNPDF2.3 LO parton distribution functions~\cite{NNPDF:2014otw}, assuming proton--proton collisions at a center-of-mass energy of $\sqrt{s} = 13.6~\mathrm{TeV}$.

At the parton level, identical baseline selection criteria are imposed on signal and background samples. Charged leptons are required to satisfy $p_\mathrm{T} > 10~\mathrm{GeV}$ and $|\eta| < 2.5$, while jets must fulfill $p_\mathrm{T} > 20~\mathrm{GeV}$ and $|\eta| < 5$. In addition, we demand that the dijet system with the largest invariant mass obey $m_{jj} > 250~\mathrm{GeV}$ in order to enhance the VBF-like topology. All cross sections are extracted directly from the \verb|MadGraph5_aMC@NLO| output and are computed at leading order in QCD.

The signal process $pp \to \chi \chi jj$ is generated through purely electroweak interactions at $\mathcal{O}(\alpha_{\rm EWK}^4)$. For the SM backgrounds, we consider single-top and $t\bar{t}$ production, Higgs production (via gluon fusion, VBF, and associated channels), vector boson production in association with jets ($V=\{W,Z\}$), QCD multijet processes, diboson production ($VV=\{WW,ZZ,WZ\}$), and triboson channels. The dominant contributions arise from $V$+jets and diboson processes with accompanying jets, particularly when gauge boson decays produce genuine missing transverse momentum, such as $Z\to\nu\nu$ and $W\to \ell\nu$. To capture these effects efficiently, we simulate the inclusive processes $pp \to \ell \nu$+jets and $pp \to \nu \nu$+jets, including both QCD-induced and purely electroweak contributions. This approach effectively accounts for the leading $V$+jets and diboson components. The combined background cross section obtained in this manner is $952.5~\mathrm{pb}$. The remaining background sources are found to be subleading after event selection and are therefore neglected in the subsequent analysis.

\begin{figure}[h!]
    \centering
    \includegraphics[width=0.235\textwidth]{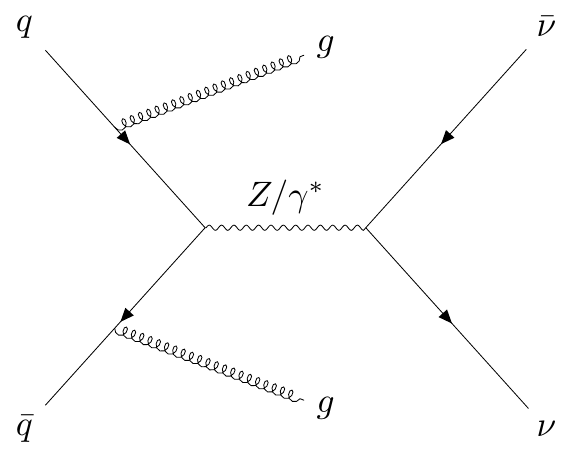}
    \includegraphics[width=0.235\textwidth]{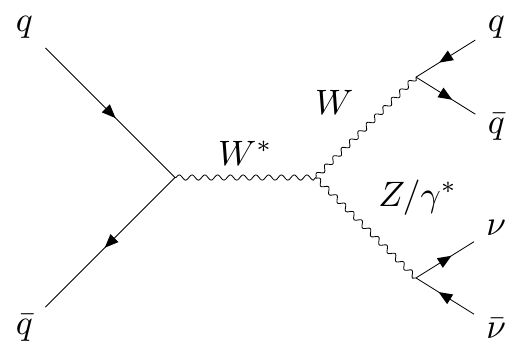}
    \caption{Representative Feynman diagrams of $pp\to \nu\bar{\nu}+$jets SM background production at the LHC.}
    \label{fig:Feyn1}
\end{figure}

\begin{figure}[h!]
    \centering
    \includegraphics[width=0.3\textwidth]{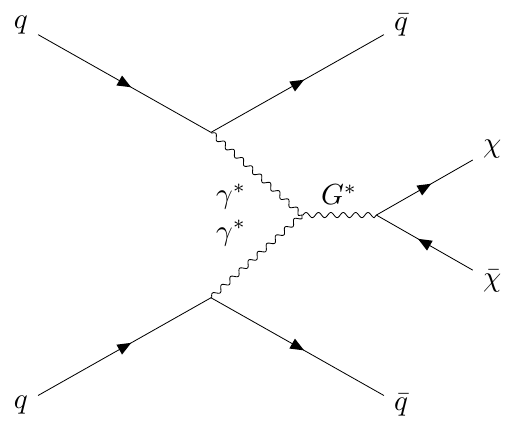}
    \caption{Representative Feynman diagram of $pp\to \chi\chi jj$ signal production via photon-photon fusion at the LHC.}
    \label{fig:Feyn2}
\end{figure}

Representative background and signal Feynman diagrams are shown in Fig.~\ref{fig:Feyn1}--~\ref{fig:Feyn2}. 
%, and the corresponding production cross section as a function of the spin-2 mediator mass $m_G$ is displayed in Fig.~\ref{fig:Cross Section}. As expected for VBF-induced production, the total rate exhibits only a moderate dependence on $m_G$, reflecting the characteristic collinear logarithmic enhancement of VBF topologies. 
Although, in principle, interference between the signal and SM background amplitudes can occur, we have explicitly verified across a broad range of $m_G$ values that such effects have a negligible impact ($\ll 1$\%) on both total rates and kinematic distributions. Signal benchmark samples are generated for mediator masses in the range $50~\mathrm{GeV} \le m_G \le 1~\mathrm{TeV}$, using variable mass steps, for fixed FIDM mass of $m_{\chi}=10$ GeV and for different benchmark choices of $\Lambda_{\gamma}$ and $\Lambda_{\chi}$.

The parton-level events are interfaced with \verb|Pythia| \allowbreak \verb|8.2.30|~\cite{Sjostrand:2014zea,Sjostrand:2007gs} for parton showering and hadronization. Detector effects are modeled using \verb|Delphes 3.4.1|~\cite{deFavereau:2013fsa}, adopting a CMS detector configuration for object reconstruction and identification. Jets are clustered with the anti-$k_\mathrm{T}$ algorithm~\cite{Cacciari_2008} as implemented in \verb|FastJet| \verb|3.4.2|~\cite{Cacciari_2012}, using a radius parameter $R=0.4$. Matching between matrix-element partons and parton-shower jets is performed using the MLM scheme~\cite{Mangano:2006rw}, with matching parameters set to \textsc{xqcut} = 30 and \textsc{qcut} = 45. These choices ensure a smooth transition in differential jet rates across different jet multiplicities.

Accurate reconstruction of leptons, light-flavor jets, and $b$-tagged jets is essential for discriminating signal events from SM backgrounds, particularly in the challenging environment expected at the HL-LHC. The large number of simultaneous proton--proton interactions per bunch crossing (pileup) degrades object identification and momentum resolution. The impact of pileup and mitigation strategies for the CMS and ATLAS experiments have been investigated in Ref.~\cite{CMS-PAS-FTR-13-014}. While a detailed simulation of the upgraded detectors is beyond the scope of this work, we incorporate conservative performance assumptions corresponding to an average of 140 pileup interactions per bunch crossing, following the guidance of Ref.~\cite{CMS-PAS-FTR-13-014}.

The reconstruction efficiency for charged hadrons, which directly affects jet and $\tau$ reconstruction as well as missing transverse momentum determination, is assumed to be approximately 97\% for $|\eta|<1.5$, decreasing to about 85\% at $|\eta|=2.5$. For electrons and muons with $p_\mathrm{T} > 5~\mathrm{GeV}$ and $|\eta|<1.5$, we assume a nominal identification efficiency of 95\% and a misidentification rate of 0.3\%~\cite{CMS-PAS-FTR-13-014,CMS:2019muj}. In the forward region $1.5<|\eta|<2.5$, these efficiencies are taken to degrade linearly, reaching 65\% with a corresponding misidentification probability of 0.5\% at the edge of the tracker acceptance. Such reductions, primarily induced by pileup, lead to a modest deterioration in the resolution of reconstructed lepton kinematics.

For $b$-jet identification, we follow Ref.~\cite{CMS:2017wtu} and adapt the \verb|Delphes| card to implement the ``Loose'' working point of the DeepCSV algorithm~\cite{Bols:2020bkb}. This choice corresponds to a $p_\mathrm{T}$-dependent $b$-tagging efficiency in the range 70--85\%, with an associated light-flavor mistag rate of approximately 10\%. The operating points for both lepton and $b$-jet identification are selected through an optimization procedure designed to maximize the expected signal significance while suppressing SM backgrounds.

\section{Machine Learning Workflow}

To discriminate signal events from Standard Model backgrounds, we employ a multivariate classification strategy based on gradient-boosted decision trees (BDTs) \cite{Friedman:2001gbm}. BDTs are a well-established machine-learning technique in high-energy physics and are particularly well suited for problems where signal and background differ through subtle, correlated features across many kinematic observables. The method constructs an ensemble of decision trees trained sequentially, where each subsequent tree is optimized to correct the misclassifications of the previous ones. The final classifier is obtained as a weighted combination of all trees, effectively performing a greedy minimization of a chosen loss function and enabling the model to learn complex, nonlinear decision boundaries.

A key advantage of BDTs is their ability to simultaneously incorporate a large number of input variables while automatically identifying the most informative combinations. This is especially powerful for collider analyses, where individual observables may offer only limited separation power, but correlated patterns across multiple kinematic quantities can provide strong discrimination. By exploiting higher-order correlations that are difficult to capture with traditional cut-based approaches, BDTs have become a standard tool in LHC analyses aimed at isolating rare signals from large backgrounds \cite{Roe:2004na,CMS:2013poe,Baldi:2014kfa,ATLAS:2017fak,Albertsson:2018maf,Barbosa:2022mmw,Qureshi:2024naw,Dutta:2022bfq}.

In this work, the machine-learning models are implemented using the \verb|Scikit-learn| and \verb|XGBoost| frameworks \cite{Pedregosa:2011ork,Chen:2016btl}. We employ the \verb|XGBClassifier| with a objective of \verb|binary:logistic| , which outputs a probabilistic estimate for signal versus background classification. The overall analysis workflow is built on a dedicated \verb|MadAnalysis| expert-mode C++ framework, which processes the simulated event samples and computes the relevant event-level and object-level observables. These quantities are exported into structured CSV files that serve as inputs to the machine-learning training pipeline. Event weights proportional to the corresponding production cross sections are applied to both signal and background samples to ensure that their relative statistical contributions are correctly represented during training.

The classifier is trained on a set of kinematic observables that characterize the dijet system and global event properties. These include the missing transverse momentum ($E_{\rm T}^{\rm miss}$), the transverse momenta of the leading and subleading jets $p_\mathrm{T}(j_1)$ and $p_\mathrm{T}(j_2)$, the dijet invariant mass $m_{jj}$, the jet pseudorapidities $\eta(j_1)$ and $\eta(j_2)$, and the angular separations $\Delta\eta$ and $\Delta\phi$ between the two jets. Jets are ordered by transverse momentum, such that $j_1$ ($j_2$) denotes the leading (subleading) jet. The full dataset, consisting of approximately $10^6$ signal events and $10^7$ background events, is randomly divided into training (90\%) and testing (10\%) subsets. Independent BDT models are trained for each assumed value of the spin-2 mediator mass.

\section{Results}\label{sec:results}

\begin{figure}[h!]
    \centering
    \includegraphics[width=0.5\textwidth]{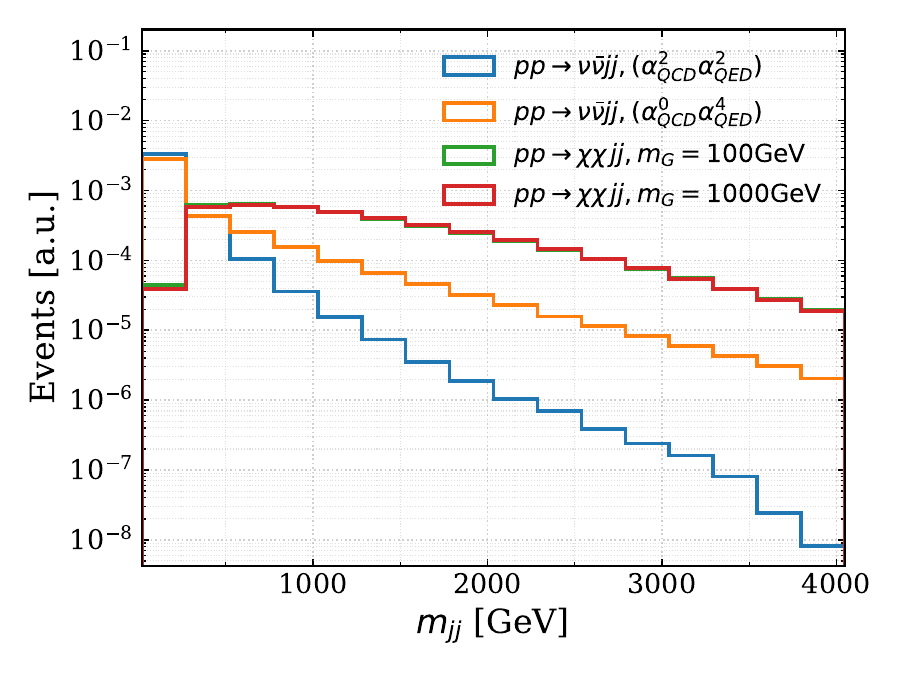}
    \caption{Dijet transverse momentum distributions for $m_G=100\text{ GeV}$ and $m_G=1000\text{ GeV}$ signals and dominant SM backgrounds}
    \label{fig:mjj}
\end{figure}

\begin{figure}[h!]
    \centering
    \includegraphics[width=0.5\textwidth]{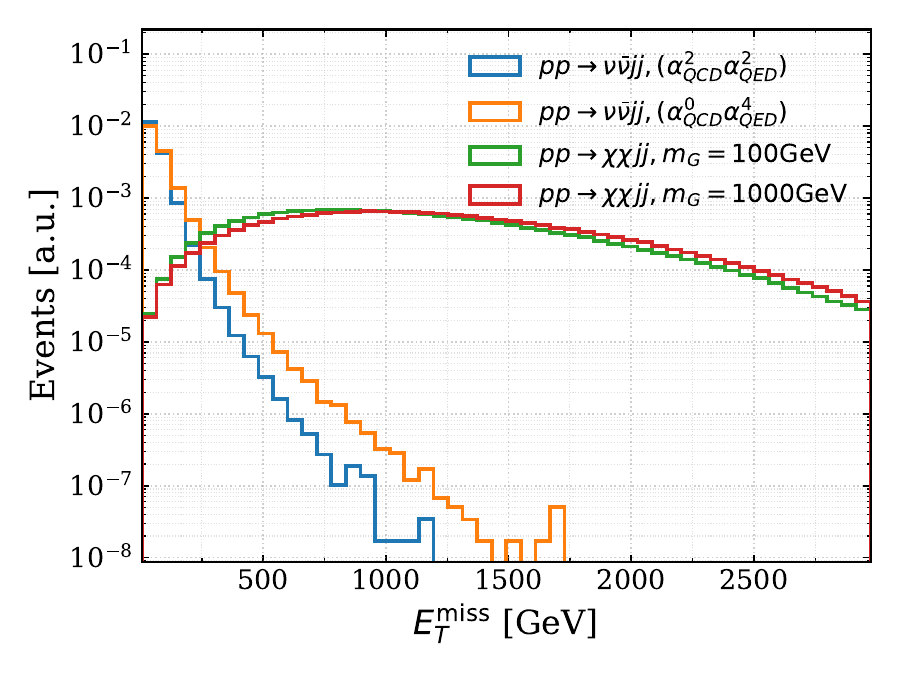}
    \caption{Missing transverse energy distributions for $m_G=100\text{ GeV}$ and $m_G=1000\text{ GeV}$ signals and dominant SM backgrounds}
    \label{fig:met}
\end{figure}

\begin{figure}[h!]
    \centering
    \includegraphics[width=0.5\textwidth]{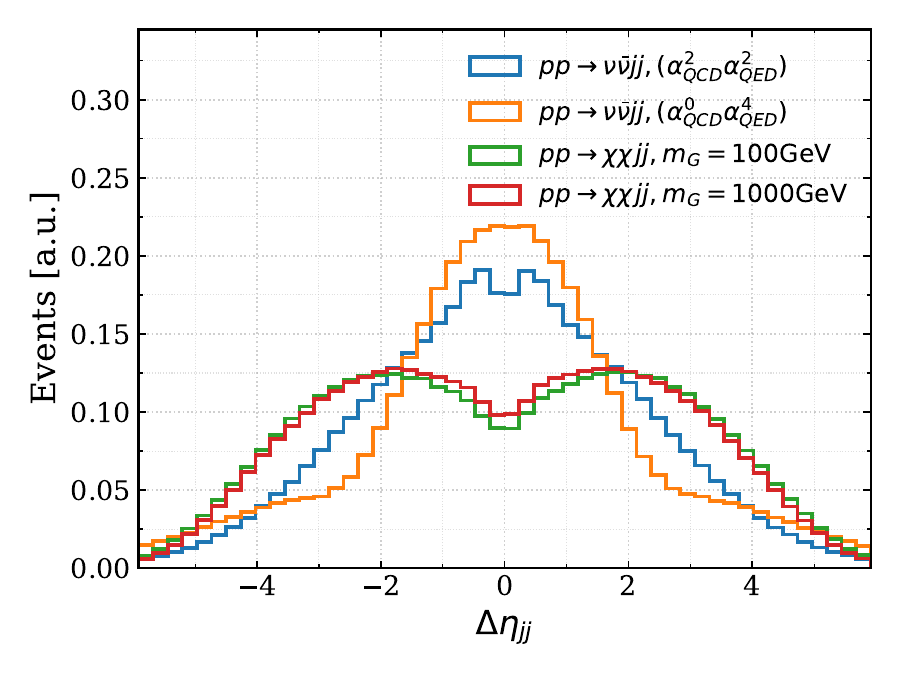}
    \caption{Pseudorapidity separation between the two jets for $m_G=100\text{ GeV}$ and $m_G=1000\text{ GeV}$ signals and dominant SM backgrounds}
    \label{fig:deltaeta}
\end{figure}

\begin{figure}[h!]
    \centering
    \includegraphics[width=0.5\textwidth]{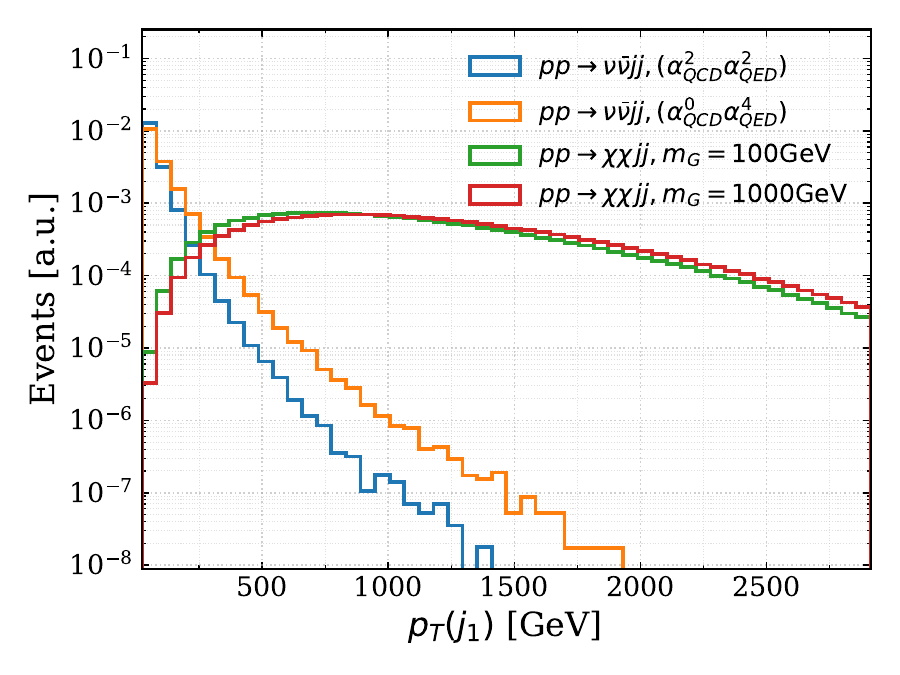}
    \caption{Transverse momentum distributions for the leading $p_T$-ordered jet for $m_G=100\text{ GeV}$ and $m_G=1000\text{ GeV}$ signals and dominant SM backgrounds}
    \label{fig:ptj1}
\end{figure}

Figures~\ref{fig:mjj}--\ref{fig:ptj1} present representative kinematic distributions for the signal benchmarks and the dominant SM backgrounds. For the pure electroweak SM background processes, the dominant contribution arises from diboson production with invisible decays ($Z\rightarrow\nu\bar{\nu}$) accompanied by hadronic gauge-boson decays ($V\rightarrow jj$), with an additional smaller contribution from electroweak VBF production of neutrinos. In contrast, SM background processes involving mixed QED--QCD interactions exhibit cross sections approximately two orders of magnitude larger, primarily due to Drell--Yan production with additional jets originating from initial-state QCD radiation. These background processes are compared to signal benchmark points corresponding to spin-2 mediator masses $m_G = 100~\mathrm{GeV}$ and $m_G = 1000~\mathrm{GeV}$.

Figure~\ref{fig:mjj} shows the reconstructed dijet invariant mass $m_{jj}$ distribution, normalized to unity. For the dominant mixed QED--QCD backgrounds, the additional jets arise predominantly from initial-state QCD radiation. Such jets typically have relatively soft transverse momenta, and their dijet invariant mass spectrum accumulates near the jet reconstruction threshold ($m_{jj}\approx p_{T1}+p_{T2}\sim 40~\mathrm{GeV}$). For the purely electroweak SM backgrounds, the jet candidates originate primarily from $V\rightarrow jj$ decays in diboson production, resulting in somewhat harder jets with characteristic scales of order $m_V$. Consequently, the corresponding $m_{jj}$ distribution is shifted to moderately larger values than the mixed QED--QCD background, but remains softer on average than the signal. 

In contrast, the signal events are produced through the VBF topology. As a result, the signal $m_{jj}$ distribution exhibits a broad spectrum with a significantly enhanced high-mass tail and a suppressed population at low $m_{jj}$, which is characteristic of VBF-like BSM processes~\cite{Shen:2026qxg,Shen:2025nkr,Qureshi:2024cmg,Gurrola:2022ssc,Cardona:2021ebw,Florez:2021zoo,Florez:2019tqr,Florez:2018ojp,Avila:2018sja,Florez:2017xhf,Florez:2016uob,Dutta:2015hra,Dutta:2014jda,Dutta:2013gga,Dutta:2012xe}. We also observe that increasing the spin-2 mediator mass produces only minimal changes in the overall $m_{jj}$ kinematic distributions. This behavior can be understood from the fact that, in VBF photon--photon fusion production at the 13.6~TeV LHC, the invariant mass of the tagging jets, $m_{jj}^{2} \simeq 2\,p_{T1}p_{T2}(\cosh\Delta\eta_{jj})$, is determined primarily by the transverse momenta and rapidity separation of the forward jets rather than by the mass of the centrally produced spin-2 mediator. The tagging jets originate from the scattered initial-state partons that radiate quasi-collinear photons, whose kinematics are governed mainly by the photon emission virtuality and the large beam energy. Consequently, for mediator masses in the range $\mathcal{O}(100~\mathrm{GeV})$--$\mathcal{O}(\mathrm{TeV})$, the characteristic jet $p_T$ spectrum and large rapidity separation remain largely unchanged, resulting in a dijet invariant mass distribution that depends only weakly on $m_G$.

Similar to current supersymmetry searches at ATLAS and CMS, the presence of DM candidates is inferred indirectly through the measurement of a momentum imbalance in the transverse plane of the detector. The reconstructed missing transverse momentum, $E_{\mathrm{T}}^{\mathrm{miss}}$, is defined as the magnitude of the negative vectorial sum of the transverse momenta of all reconstructed visible particle candidates, as shown in Figure~\ref{fig:met}. In the dominant SM backgrounds, the missing transverse momentum originates primarily from neutrinos produced in $W$ or $Z$ boson decays, which typically carry modest transverse momentum due to the electroweak mass scale $m_V$ and the relatively balanced kinematics of the underlying hard-scattering process. Consequently, the background distributions peak at relatively low $E_{\mathrm{T}}^{\mathrm{miss}}$ values. In contrast, signal events produced via the VBF mechanism tend to exhibit significantly larger missing transverse momentum due to the higher mass scale $m_G$ associated with the mediator and the recoil of the invisible DM system against the forward tagging jets. This results in a harder $E_{\mathrm{T}}^{\mathrm{miss}}$ spectrum with an extended high-$E_{\mathrm{T}}^{\mathrm{miss}}$ tail.

Figure~\ref{fig:deltaeta} shows the pseudorapidity separation between the two leading jets, $\Delta\eta_{jj}$. A defining feature of the VBF topology is the presence of two forward jets with large rapidity separation, which leads to signal events populating the larger $|\Delta\eta_{jj}|$ region. In contrast, mixed QED-QCD backgrounds such as Drell--Yan plus jets production tend to produce jets more centrally, resulting in a distribution that peaks at small pseudorapidity separation. The pure electroweak background contains contributions from both diboson production with relatively central jets and electroweak VBF neutrino production, leading to a small but visible population extending into the larger $|\Delta\eta_{jj}|$ region.

Finally, Figure~\ref{fig:ptj1} shows the transverse momentum distribution of the leading jet, $p_T(j_1)$, for the dominant backgrounds and signal benchmarks. In background processes, jets are frequently produced through initial-state radiation or from relatively low transverse momentum boson recoil, leading to a spectrum dominated by softer jets. In contrast, the VBF production mechanism in the signal generates two energetic forward jets originating from the scattered quarks, resulting in a harder leading-jet $p_T$ spectrum relative to the backgrounds. This harder jet activity, together with the large rapidity separation and enhanced dijet invariant mass, provides a characteristic signature that helps discriminate the VBF signal from the SM backgrounds.

\begin{figure}[h!]
    \centering
    \includegraphics[width=0.5\textwidth]{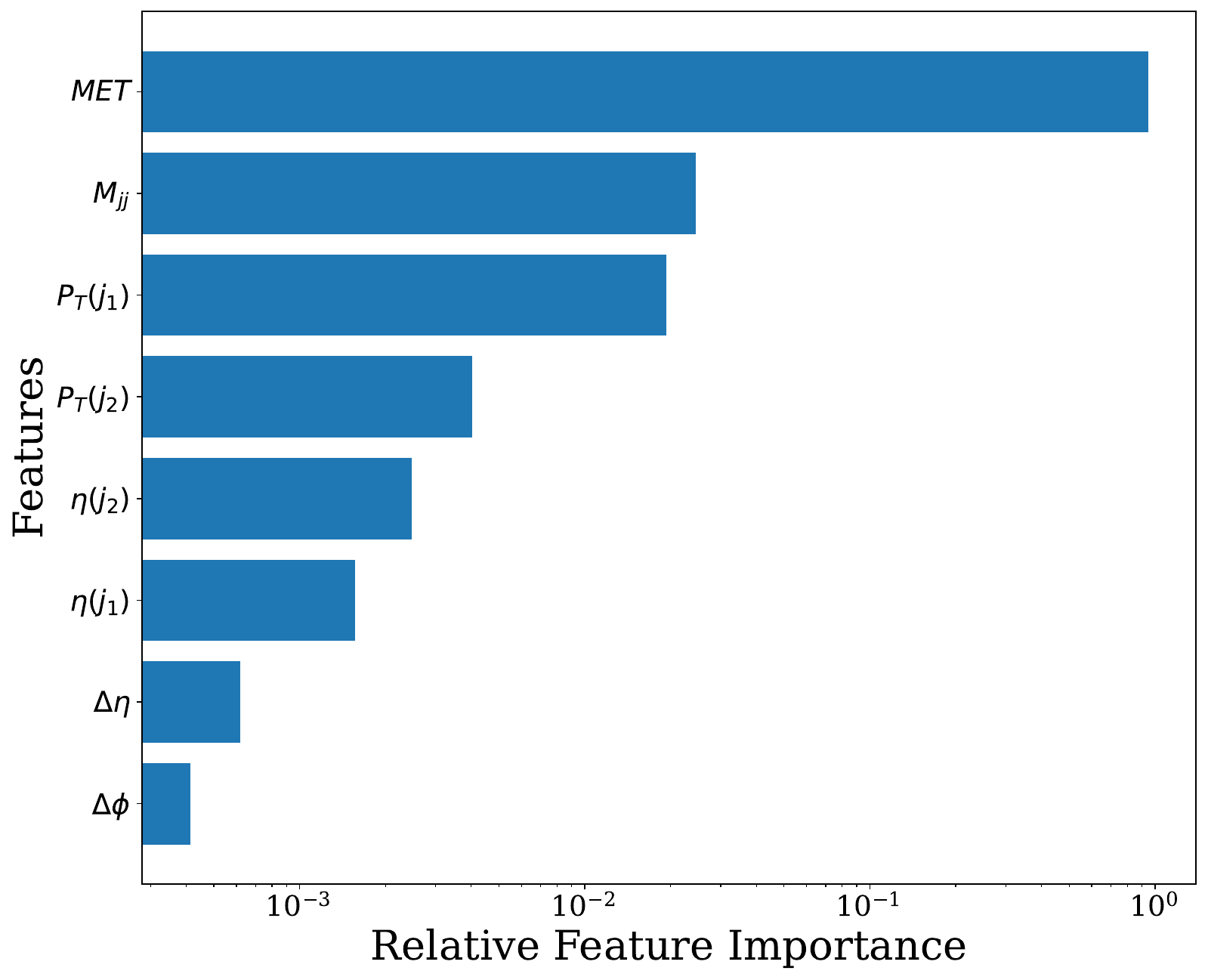}
    \caption{Relative importance of features in training for a benchmark with $m_G = 1000 \text{ GeV}$ and $m_\chi = 10 \text{ GeV}$}
    \label{fig:featimp}
\end{figure}

In addition to these aforementioned variables in Figures~\ref{fig:mjj}-\ref{fig:ptj1}, 
several other kinematic variables were included as inputs to the BDT algorithm. As described previously, 8 such variables were used in total, which included the momenta and pseudorapidities of jets candidates, the missing transverse momentum, the invariant mass of pairs of jets, and angular difference $\Delta\eta(jj)$ and $\Delta\phi(jj)$ between the jet pair. The BDT learning model returns the discriminating power of each of its inputs, and the feature importance for each variable is shown in Figure~\ref{fig:featimp} for a signal benchmark point with $m_G = 1000 \text{ GeV}$, $\Lambda_{\gamma}=\Lambda_{\chi}=1$ TeV.

\begin{figure}[h!]
    \centering
    \includegraphics[width=0.5\textwidth]{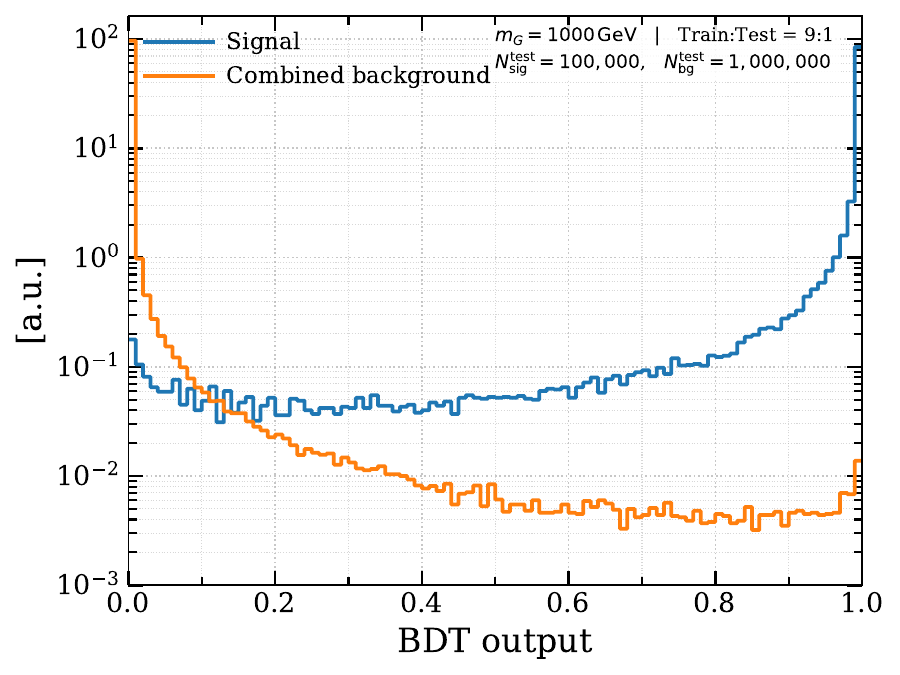}
    \caption{BDT output distributions for a benchmark with $m_G = 1000 \text{ GeV}$ and $m_\chi = 10 \text{ GeV}$. The distributions are normalized to unity}
    \label{fig:BDToutput}
\end{figure}

The separation between signal and background is achieved using the BDT classifier described in the previous section. The resulting BDT response distributions for representative signal hypotheses, $m_G = 1000$~GeV and $m_G = 100$~GeV, are shown in Fig.~\ref{fig:BDToutput}, together with the dominant SM backgrounds. For visualization purposes, these distributions are normalized to arbitrary units such that the total area of the signal and background histograms is equal. The BDT output score ranges between 0 and 1, with values approaching unity corresponding to increasingly signal-like events, while values near zero indicate background-like behavior.

For the sensitivity projections, the BDT output distributions are instead normalized to the expected event yields according to $N = \mathcal{L}\,\sigma\,\epsilon$, where $\sigma$ denotes the signal or background cross section, $\epsilon$ represents the overall selection efficiency (including reconstruction and identification effects), and $\mathcal{L}$ is the integrated luminosity. Throughout this analysis, we assume $\mathcal{L} = 3000~\text{fb}^{-1}$, corresponding to the expected dataset of the HL-LHC. The full BDT output spectrum is used in a shape-based analysis, in which the binned distributions are incorporated into a profile likelihood fit to extract the expected $95\%$ confidence level upper limits on the signal cross section. The statistical interpretation is performed using the \texttt{Combine} framework~\cite{CMS:2024onh}, employing the asymptotic approximation with an Asimov dataset. This approach allows us to exploit the full discriminating power of the BDT output rather than relying on a single optimized selection.

Systematic and statistical uncertainties are incorporated into the likelihood model via nuisance parameters. Normalization uncertainties are implemented using log-normal priors, while shape-dependent uncertainties are modeled with Gaussian constraints affecting the bin-by-bin yields of the BDT distributions. Both theoretical and experimental sources of uncertainty are taken into account in a consistent manner.

Theoretical uncertainties associated with parton distribution functions (PDFs) are evaluated following the PDF4LHC recommendations~\cite{Butterworth:2015oua}. These uncertainties primarily affect the overall normalization of the signal and background yields, with negligible impact on the shape of the BDT output distributions compared to statistical fluctuations. Accordingly, PDF uncertainties are treated as uncorrelated between signal and background processes, but fully correlated across BDT bins for a given process. Additional theoretical uncertainties arise from missing higher-order corrections in the signal prediction. These are estimated by varying the renormalization and factorization scales by a factor of two up and down from their nominal values and propagating the resulting variations to the BDT distributions. The corresponding uncertainties are found to be at the level of $1$--$3\%$, depending on the mediator mass and the region of the BDT output.

Experimental uncertainties are also included following current LHC measurements. A $3\%$ uncertainty on the integrated luminosity is applied and treated as fully correlated across all processes and bins~\cite{lumiRef}. For leptons, we assign a $2\%$ uncertainty related to reconstruction, identification, and isolation efficiencies, along with an additional $3\%$ uncertainty to account for energy/momentum scale and resolution effects, assumed to be independent of $p_T$ and $\eta$. These uncertainties are taken as correlated across processes containing genuine leptons and across BDT bins. Jet-related uncertainties are incorporated through variations of the jet energy scale, typically in the range of $2$--$5\%$ depending on $p_T$ and $\eta$~\cite{EXO19015,EXO17016,EXO16016}, which induce shape variations in the BDT output at the level of $1$--$3\%$ across bins. 

Since an assessment of a detailed data-driven background estimation strategy is beyond the scope of this work, we include an additional conservative normalization uncertainty of $10\%$ to account for potential mismodeling in both signal and background predictions. This uncertainty is treated as uncorrelated between different processes and independent of the BDT output bin. Combining all sources, the total systematic uncertainty is estimated to be of order $20\%$.

In addition to the above experimental uncertainties, we also take into account statistical uncertainties arising from the number of simulated signal and background events that pass the selections and fall into a specific bin of the BDT output histogram. These uncertainties vary depending on the process and the particular BDT output bin. These uncertainties are uncorrelated across processes and uncorrelated across BDT output bins.

\begin{figure}[h!]
    \centering
    \includegraphics[width=0.5\textwidth]{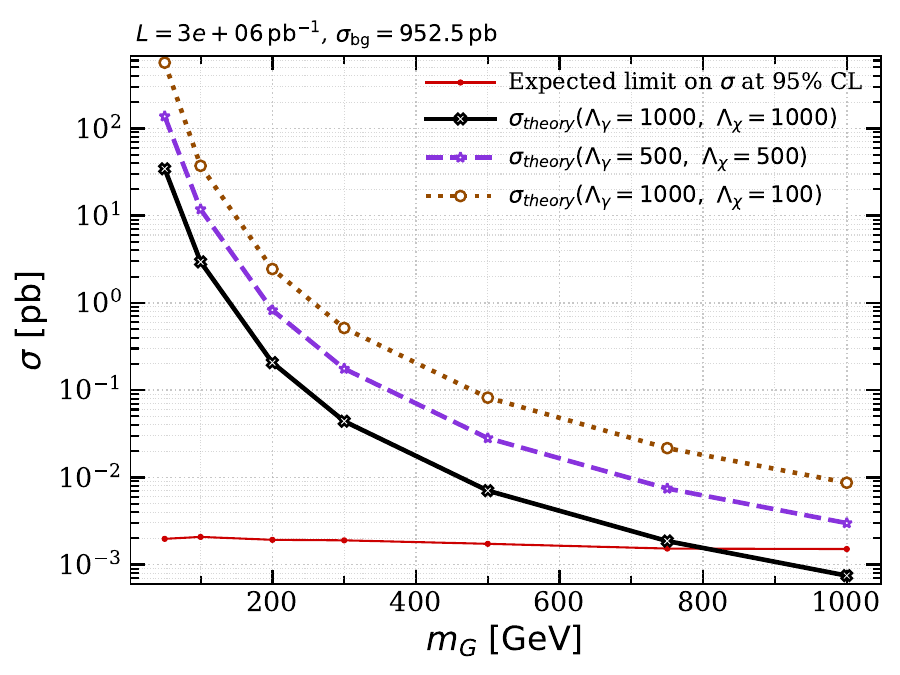}
    \caption{Projected 95\% CL upper limits on the production cross section $\sigma(pp \to Gjj \to \chi\chi jj)$ for a range of mediator mass from $m_G=50~\mathrm{GeV}$ to $m_G=1000~\mathrm{GeV}$ at the HL-LHC with $L = 3000~\mathrm{fb}^{-1}$. The expected experimental sensitivity is compared to theoretical predictions for representative coupling choices.}
    \label{fig:crossSecExcl}
\end{figure}

\begin{figure*}
\centering
\begin{minipage}{.5\textwidth}
  \includegraphics[width=\linewidth]{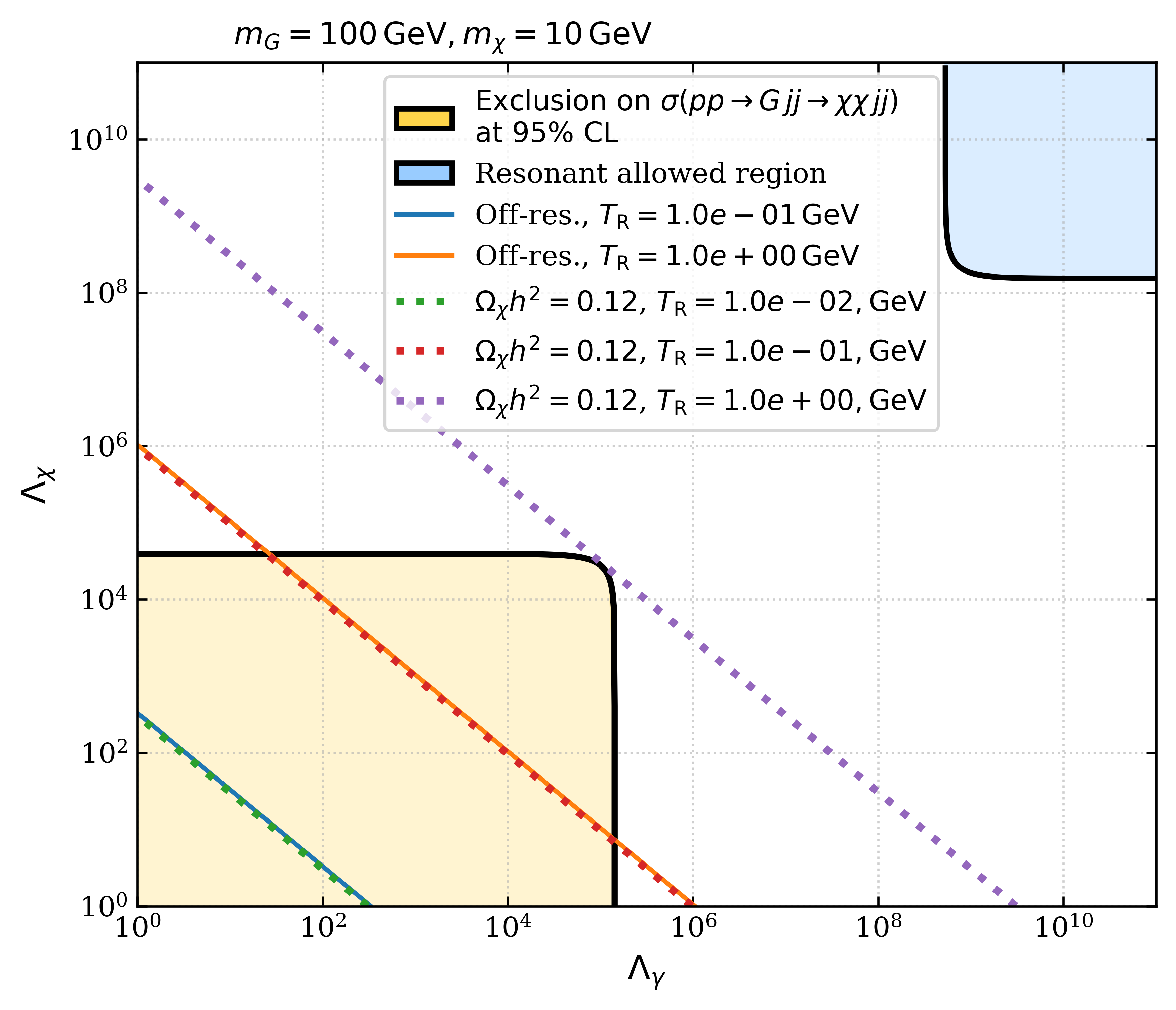}
\end{minipage}%
\begin{minipage}{.5\textwidth}
  \centering
  \includegraphics[width=\linewidth]{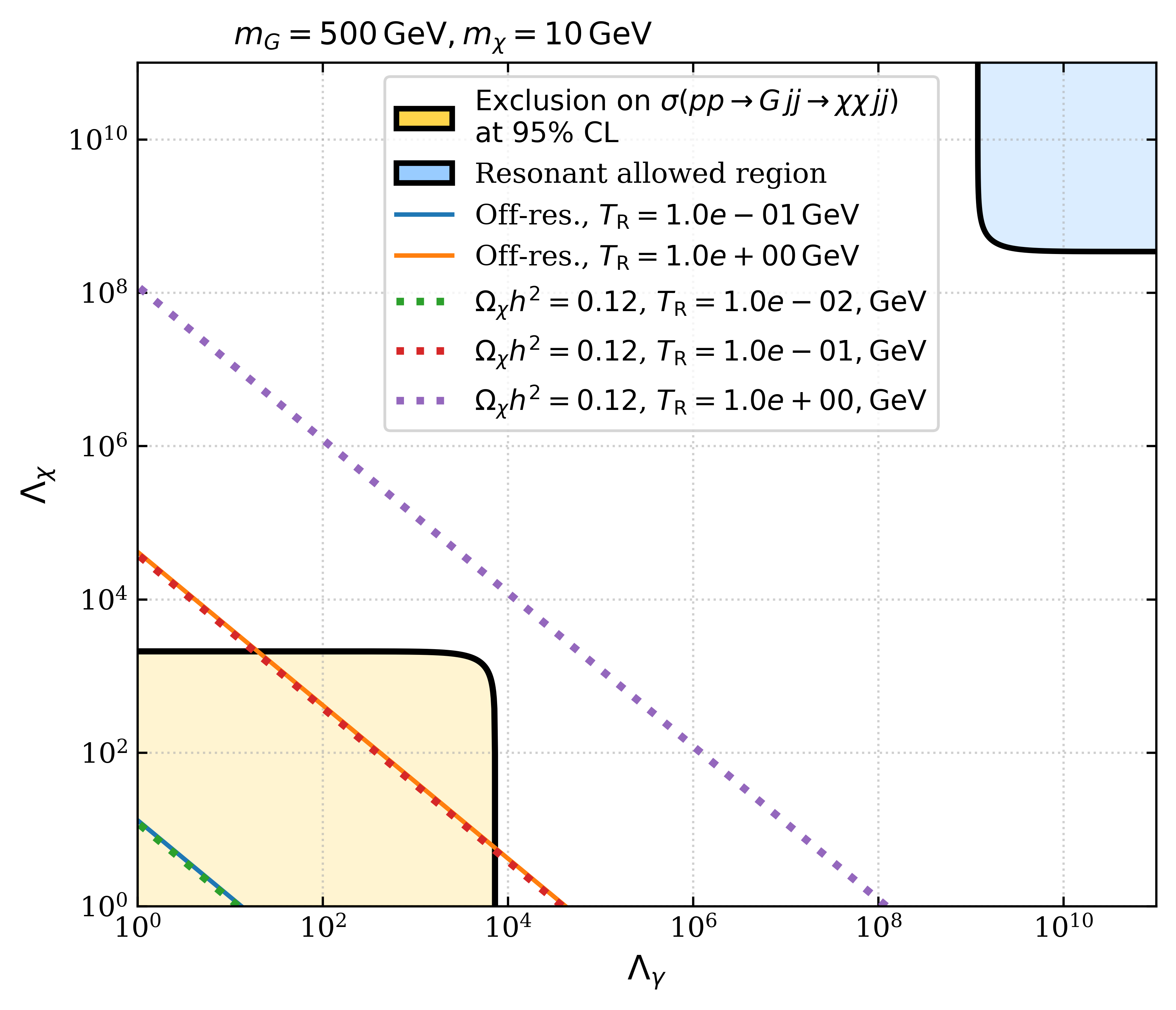}
\end{minipage}%

\begin{minipage}{.5\textwidth}
  \includegraphics[width=\linewidth]{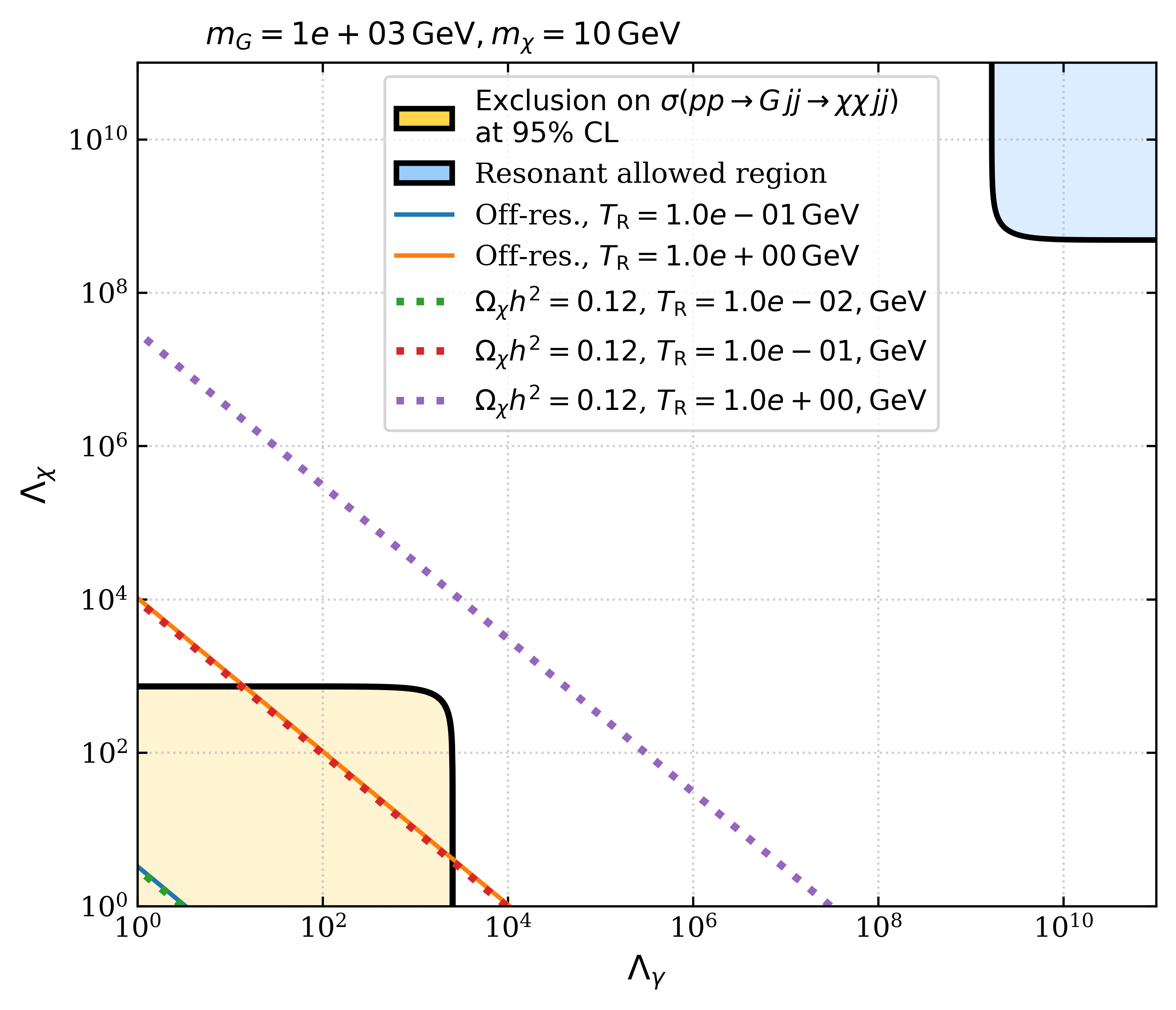}
\end{minipage}%
    \caption{Projected constraints on the effective couplings $\Lambda_\gamma$ and $\Lambda_\chi$ for dark matter mass $m_\chi = 10~\mathrm{GeV}$ and a spin-2 mediator with $m_G = 100~\mathrm{GeV}$, $m_G = 500~\mathrm{GeV}$, and $m_G = 1~\mathrm{TeV}$ at the HL-LHC. The yellow shaded region indicates the $95\%$ CL exclusion on $pp \to Gjj \to \chi\chi jj$ using BDT workflow, while dotted contours show parameter space consistent with the observed dark matter relic abundance and off-resonance dominated production for different reheating temperatures.}
    \label{fig:SSLS3000}
\end{figure*}

We compare the projected 95\% CL limits on the signal cross section obtained from the profile likelihood analysis with the theoretical signal production cross section times branching fraction for different coupling combinations. 
%and examine the corresponding cosmological constraints. For the following results, we assume an integrated luminosity of $3000~\mathrm{fb}^{-1}$, representative of the full High-Luminosity LHC dataset, and a background systematic uncertainty of 1\%. 
Figure~\ref{fig:crossSecExcl} shows the projected 95\% CL exclusion cross section overlaid with the theoretical cross section for three representative coupling choices, across the mediator mass range $m_G = 50$–$1000$ GeV. 
%For a given coupling combination, the production cross section is obtained by rescaling a reference simulated cross section by $1/\Lambda_\gamma^2$, and multiplying by the corresponding branching ratio evaluated at that coupling point (see Eq.~??). 
For $(\Lambda_\chi, \Lambda_\gamma) = (1000, 1000)$, the expected exclusion reach extends to approximately $m_G \sim 800$ GeV at 95\% CL. For smaller $\Lambda$ scales (corresponding to larger production rates), TeV-scale mediator masses become excludable. Due to the narrow width of the spin-2 mediator, the event kinematics—and therefore the BDT response—do not significantly change at larger $m_G$, implying that the exclusion cross section above 1 TeV remains approximately $2 \times 10^{-3}$ pb within this framework.

To evaluate cosmological consistency, we construct phase-space plots in the $(\Lambda_\chi, \Lambda_\gamma)$ plane. We overlay the off-resonance freeze-in condition (Eq.~\ref{eq:freezeInCondition}), the resonance freeze-in condition ($\Gamma_{G} < H(T)$), and the relic density requirement (Eq.~\ref{eq:relicDensity}), imposing that the predicted dark matter abundance satisfies $\Omega h^2 \le 0.12$, consistent with cosmological observations. The resonance region corresponds to very large coupling scales and suppressed production rates. For the off-resonance freeze-in and relic density constraints, three reheating temperatures are considered: $T_R = 10$ MeV, 100 MeV, and 1 GeV. This range is motivated by studies suggesting that the reheating temperature in the early universe may lie at or below the MeV scale~\cite{Kawasaki:1999na,Hannestad:2004px,deSalas:2015glj}.

The collider exclusion region is determined by identifying all coupling combinations that yield a production cross section exceeding the BDT-derived 95\% CL limit at a given $m_G$. We consider three benchmark mediator masses: 100 GeV, 500 GeV, and 1 TeV, as shown in Fig.~\ref{fig:SSLS3000}. For the low-, intermediate-, and high-mass cases, the exclusion reach extends approximately to $(\Lambda_\chi, \Lambda_\gamma) \sim (10^4, 10^5)$, $(10^3, 10^4)$, and $(10^3, 10^3)$, respectively. In all three mass scenarios, the relic density constraint at $T_R = 10$ MeV and the off-resonance freeze-in constraint at $T_R = 100$ MeV lie entirely within the collider-excluded region. For $T_R = 100$ MeV relic density and $T_R = 1$ GeV off-resonance freeze-in, a portion of the cosmologically viable parameter space—particularly where $\Lambda_\chi$ and $\Lambda_\gamma$ are of comparable magnitude—is excluded. However, at $T_R = 1$ GeV, the region allowed by the relic density constraint remains outside the collider exclusion reach for all three mediator masses.

\section{Discussion}\label{sec:discussion}
In this work, we have investigated the sensitivity of vector boson fusion (VBF) topologies, combined with a boosted decision tree (BDT) classifier, to probe the electroweak production of a spin-2 mediator coupled to photons and feebly interacting dark matter. Our results demonstrate that the inclusion of machine learning techniques substantially enhances signal--background discrimination compared to traditional cut-based strategies, thereby extending the reach of LHC searches into previously unexplored regions of parameter space. Assuming an integrated luminosity of $\mathcal{L} = 3000~\mathrm{fb}^{-1}$ and realistic systematic uncertainties expected at the High-Luminosity LHC (HL-LHC), mediator masses approaching the TeV scale become accessible for moderate values of the effective coupling scales.

From an experimental standpoint, the VBF topology is essential in achieving this sensitivity. The presence of two forward jets with large invariant mass and significant rapidity separation provides powerful suppression of Standard Model backgrounds. In addition, the missing transverse momentum associated with dark matter production offers a complementary handle for signal identification. The BDT classifier further improves the analysis by exploiting non-trivial correlations among multiple kinematic observables, enabling a more efficient separation of signal and background than is achievable with simple rectangular cuts.

To place our results in context, we compare with existing LHC diphoton resonance searches. These analyses are typically interpreted within models where the spin-2 mediator couples strongly to gluons and quarks, leading to significantly enhanced production cross sections. As a result, current experimental limits are derived for cross sections that are orders of magnitude larger than those predicted in our photon-coupled, feebly interacting scenario. In the parameter space relevant for this work, where the signal cross section is typically $\mathcal{O}(10^{-3})$~pb or smaller, inclusive diphoton searches have negligible sensitivity. This highlights the importance of developing dedicated search strategies tailored to weakly coupled scenarios.

We quantify the collider reach by comparing the expected $95\%$ CL upper limits obtained from the BDT-based analysis with the predicted signal cross sections across representative coupling choices. Within the weakly coupled regime, the HL-LHC sensitivity extends to mediator masses approaching the TeV scale, with the exclusion threshold reaching the level of a few $\times 10^{-3}$~pb at high $m_G$. The relatively mild dependence of the sensitivity on $m_G$ reflects the stability of the VBF kinematics and the corresponding BDT response as the mediator mass increases.

Finally, we assess the interplay between collider constraints and cosmological requirements by overlaying the projected exclusion contours with the regions of parameter space consistent with freeze-in production and the observed relic abundance in the $(\Lambda_\chi, \Lambda_\gamma)$ plane. We find that collider searches at the HL-LHC can probe a significant fraction of the parameter space compatible with low reheating temperature scenarios, where off-resonant freeze-in dominates. In contrast, scenarios with higher reheating temperatures, particularly those involving resonant production, typically require much larger coupling scales and remain largely beyond the reach of the HL-LHC.

Overall, this work establishes a coherent framework connecting early-Universe cosmology with collider phenomenology in the context of spin-2 mediated freeze-in dark matter. It demonstrates that VBF-based searches augmented by machine learning techniques provide a viable and powerful strategy to explore feebly interacting dark sectors at the HL-LHC, and it motivates further experimental and theoretical efforts to refine and extend these approaches.

\noindent \textbf{Acknowledgements:}A. G and J. L. acknowledge the funding received from the Physics \& Astronomy department at Vanderbilt University and the US National Science Foundation. This work is supported in part by NSF Awards PHY-1945366 and PHY-2411502.

\bibliographystyle{spphys}      
\bibliography{refs.bib}   

\end{document}